\documentclass{aa}
\usepackage{graphicx,natbib}
\def\hra{HR\,4796\,A}
\def\hrb{HR\,4796\,B}
\def\hda{HD\,\-141569\,A}
\def\hdb{HD\,141569\,B}
\def\hdc{HD\,141569\,C}
\def\psf{HD\,145788}
\def\cs{\mbox{circumstellar}}
\def\co{\mbox{coronagraphic}}

\begin{document}
\thesaurus{08 (19.42.1 \hda\ ; 19.25.1; 04.01.1; 05.01.1; 19.94.1)}
           %
           %
\authorrunning{J.C. Augereau et al.}  \titlerunning{HST/NICMOS2
  observations of the \hda\ \cs\ disk}
\title{ HST/NICMOS2
  observations of the \hda\ \cs\ disk
\thanks{Based on observations
    with the NASA/ESA Hubble Space Telescope, obtained at the Space
    Telescope Science Institute, which is operated by the Association
    of Universities for Research in Astronomy, Inc. under NASA
    contract No. NAS5-26555.}}
\author{J.C. Augereau\inst{1} \and A.M. Lagrange\inst{1} \and
D. Mouillet\inst{1} \and F. M\'enard\inst{1,2}}
\offprints{J.C. Augereau}
\mail{augereau@obs.ujf-grenoble.fr}
\institute{
Laboratoire d'Astrophysique de l'Observatoire de Grenoble, Universit\'e J.
Fourier / CNRS, B.P. 53, F-38041 Grenoble Cedex 9, France
\and Canada-France-Hawaii Telescope Corporation, PO Box 1597, Kamuela,
HI 96743, USA}
\date{Received {}; accepted {}}
\maketitle
\begin{abstract}

  We report the first resolved scattered light images of the \cs\ dust
  disk around the old Pre-Main Sequence star \hda.  The disk seen in
  HST/NICMOS2 images shapes a bright annulus inclined at $37.5\degr\pm
  4.5\degr$ from edge-on.  This ring peaks at $325\pm10$\,AU from the
  star with a characteristic width of $\sim 150$\,AU. At $1.6\,\mu$m,
  the dust grains scatter a total flux density of at least $4.5\pm
  0.5$\,mJy.  Our disk model using the spatial distribution implied by
  the images does not explain the $10\,\mu$m excess and requires an
  additional grain population closer to the star. Some similarities
  and differences with the dust annulus surrounding \hra\ are pointed
  out.  
\end{abstract}
%
%
\section{Introduction}
The recent discovery of a disk around the old Pre-Main Sequence A0
star \hra\ \citep{koe98,jaya98,aug99,sch99} opened new perspectives in
our understanding of the evolution of circumstellar disks and the
early stages of planetary formation \citep{lag99}.  The comparable
age, spectral type and IRAS infrared excess of the post Herbig Ae/Be
star \hda\ are pertinent clues for suspecting the presence of an
optically thin disk around this star at a similar evolutionary status
($t_*\geq\,10\,$Myr, B9.5Ve star, van den Ancker et al. 1998).

A main difference with \hra\ may concern the multiplicity.  Whereas
\hra\ has a physically bounded companion (\hrb), which may play a role
in the dynamics of the disk, the two stellar companions of \hda\ 
detected so far \citep{gahm83,pir97} may not be gravitationally linked
to the primary, as postulated by \citet{lin85}. In addition, no
spectroscopic companion is detected by \citet{cor99}.

Since IRAS, further investigations of the material around \hda\ have
been performed. \hda\ shows a very small intrinsic polarization
consistent with that of the prototype Vega-like stars
\citep{yud98,yud99}.  Emission spectral features from \cs\ dust grains
have also been observed by \citet{syl96a} at 10--20\,$\mu$m.

Whereas the spectral energy distribution (SED) constrains the grain
composition, it poorly constrains the shape of the dust distribution.
In particular, models predict an inner edge for the disk
ranging between 10\,AU \citep{mal98} and 650\,AU \citep{syl96b}.

In this Letter, we present the first resolved scattered light images
of the \hda\ \cs\ disk obtained with the coronagraph on the
HST/NICMOS2 camera. A disk detection is also reported independently by
\citet{wei99} at $1.1\,\mu$m.  We afterwards detail the morphology of
the resolved structure, its brightness at $1.6\,\mu$m and finally
discuss some disk properties.
%
\section{Observations and reduction procedure}
\subsection{The data}
HST/NICMOS2 \co\ observations of \hda\ ($V=7.1$) were obtained on 1998
August 17 in MULTIACCUM mode. Six consecutive exposures on the target
were performed in filter F160W (central wavelength $\simeq 1.6\,\mu$m,
bandwidth $\simeq 0.4\,\mu$m) corresponding to a total integration
time of 14\,m\,23\,s.  The reduction procedure of \co\ data requires a
comparison star to assess the Point Spread Function (PSF).  For this,
the A1V star \psf, which shows no evidence of \cs\ material, was
observed during the same orbit for 6\,m\,24\,s to achieve a similar
signal to noise ratio given its own flux ($V=6.25$).

\begin{figure}[tbp]
\begin{center}
\includegraphics[angle=90,origin=bl,width=0.49\textwidth]{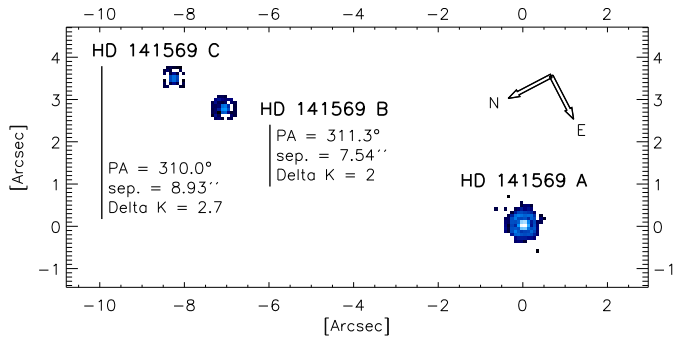}
\caption[]{The multiple system HD\,141569 in filter F171M (central
  wavelength $\simeq 1.72\,\mu$m, bandwidth $\simeq 0.07\,\mu$m) and
  known astronomical parameters for the companions \hdb\ and \hdc. K
  magnitude differences are from \citet{pir97}.}
\label{field}
\end{center}
\end{figure}

A narrow band (filter F171M) view of the field around \hda\ taken
during the target acquisition confirms the presence of two bright
companions \hdb\ and \hdc\ previously identified in K band by
\citet{pir97} (Fig.\ref{field}).  Measured position angles (hereafter
PA) and separations from the primary star are summarized in Figure
\ref{field}. PAs are fully consistent with \citet{pir97} results
whereas distances from this work are about 11\% larger than those
measured by these last authors.
\begin{figure*}[tbp]
\begin{center}
\includegraphics[angle=90,origin=bl,width=0.95\textwidth]{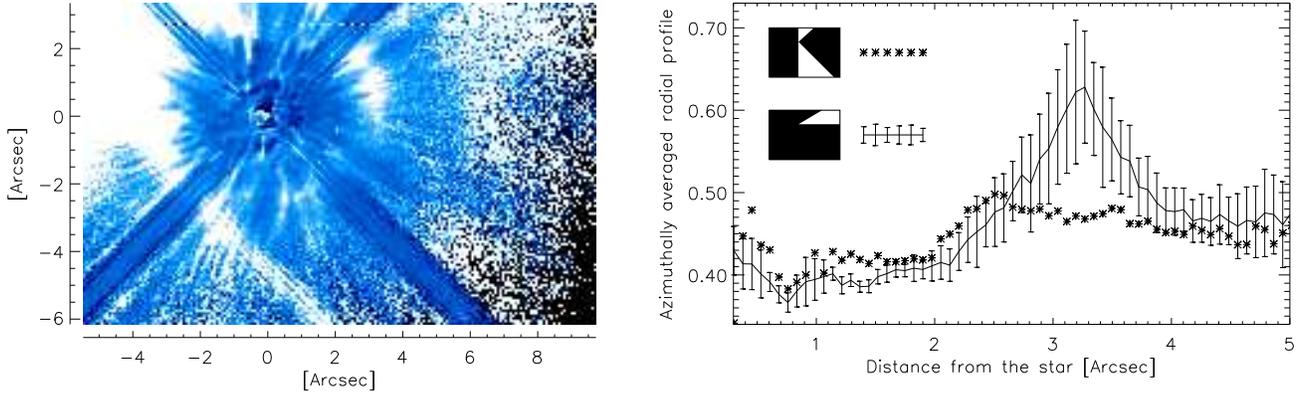}
\caption[]{\textit{Left}: ratio of the \hda\ image to the reference star
  image assumed to be representative of the PSF. The image reveals an
  excess corresponding to the presence of \cs\ material well centered
  on the star.  The top-left corner is blurred by the companions \hdb\ 
  and \hdc.  The mask has a radius of 0.3$\arcsec$ ($\sim$\,4 pixels;
  the pixel size is about 0.076$\arcsec$). \textit{Right}: two
  azimuthally averaged radial profiles.  The small rectangular boxes
  mimic the left-hand image, with the white angular sectors
  illustrating the areas used to compute the profiles.  Between
  $0.8\arcsec$ and $2\arcsec$, the signal to noise ratio in the
  profile plotted as asterisks is roughly a half the signal to noise
  ratio for the other profile.}
\label{div}
\end{center}
\end{figure*}
\subsection{Reduction procedure for the \co\ data~: basic cleaning and
PSF subtraction}
For both \hda\ and the PSF reference (\psf), we added the calibrated
files provided by the STScI. We cleaned the bad pixels and 'grots'
\citep{stsci} using {\sc Eclipse} reduction procedures
\citep{eclipse}. Blurred stripes on the images associated to
electronic echos of the source (also called 'Mr Staypuft' ghosts,
STScI 1997) were subtracted by averaging a profile perpendicular to
the stripes in a region free of astronomical sources.

Before being subtracted, the reference star flux has to be scaled to
that of the target object.  The ratio of the \hda\ image to the
reference star image gives the scaling factor. At this stage of the
reduction, this is also an {\it unbiased} and powerful way to detect
\cs\ material.  Indeed, a \cs\ structure is expected to appear as a
continuous feature in the ratio at a level significantly higher than
the background level.

Figure \ref{div}\,(\textit{left}) shows this ratio. An annular
structure centered on the star clearly appears especially in the right
and bottom parts of the image whereas the light of the stellar
companions (mainly \hdb) contaminates the opposite side.  Azimuthally
averaged radial profiles on different areas of the ratio confirm the
presence of an excess (Fig.\ref{div},\,\textit{right}).

The true linear resolution is higher in an angular sector close to the
major axis of the annular structure than in the perpendicular
direction (resp. solid line and bold asterisks
Fig.\ref{div}, \textit{right}).  The superimposition of the two
azimuthally averaged radial profiles shows that these profiles have a
similar behaviour and are roughly a constant up to $\sim 2\arcsec$,
then shows a strong discrepancy between $2\arcsec$ and $4\arcsec$. We
assume then that the region up to $2\arcsec$ is free of significant
resolved dust amount and that the scaling factor is $\sim 0.400\pm
0.015$ (between $0.8\arcsec$ and $2\arcsec$).
\section{Results}
\begin{figure*}[tbp]
\begin{center}
\includegraphics[angle=90,origin=bl,width=0.99\textwidth]{bg282.f3}
\caption[]{\textit{Left}:  Scattered light image ($1.6\,\mu$m) of the \hda\
  \cs\ disk in logarithmic scale. The scattered light due to the
  companions \hdb\ and \hdc\ and to their associated spider
  diffraction spikes (see Fig.\ref{div} ) have been subtracted. In the
  bottom-right corner, we show the same disk where most of the
  unrealistic bright patterns close to the star (see text) have been
  removed so as to highlight the annular resolved structure.
  \textit{Right}: Contours of the disk to allow comparison with the
  ellipse fitting superimposed on the image.}
\label{disk}
\end{center}
\end{figure*}
\subsection{Orientation of the disk}
Figure \ref{disk}\,(\textit{left}) shows the final reduced image of
the disk and brings out the annular structure evidenced in the ratio
of \hda\ to the reference star (Fig.\ref{div}).  Unsubtracted
secondary spider diffraction spikes are responsible for the bright
areas inside the annulus. These patterns do not correspond to any
realistic excess. An image of what would be observed without these
spikes is shown in the bottom-right corner of the reduced image.

We computed the distance from the star corresponding to the maximum
surface brightness of the annulus versus the position angle over a
90$\degr$ range (Southern part of the disk). Least-squares ellipse
fitting constrains the major axis of the observed annulus to be at a
position angle of $355.4\degr\pm 1\degr$ and leads to an upper limit
for the disk inclination from edge-on of $37.5\degr\pm 4.5\degr$
assuming that the disk is axisymmetrical with respect to the star. The
fit is superimposed on the reduced image of the disk
(Fig.\ref{disk},\,\textit{right}) and agrees well with all the
observed structure.
\subsection{Surface brightness distribution and photometry}
\begin{figure}[tbp]
\begin{center}
\includegraphics[angle=90,origin=bl,width=0.5\textwidth]{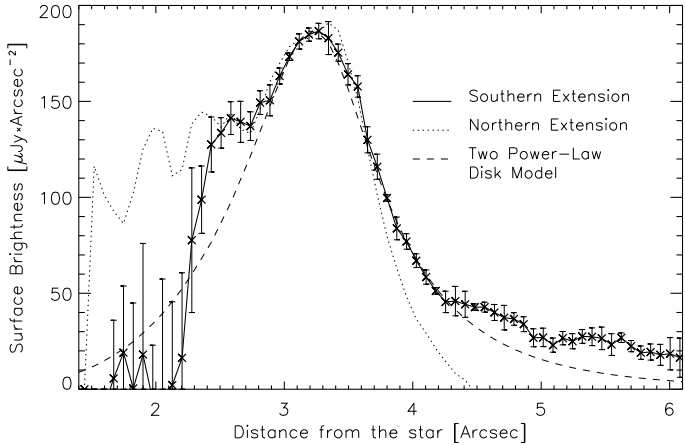}
\caption[]{Radial surface brightness distribution along the major axis of the
  disk. They have been obtained by azimuthally averaging the profiles
  over a 6 degree sector overlapping the major axis.}
\label{profile}
\end{center}
\end{figure}

Both radial surface brightness profiles along the Northern and the
Southern semi-major axis of the disk peak at $325\pm 10$\,AU from the
star, according to the \textit{Hipparcos} star distance of
$99^{+9}_{-8}$\,pc (Fig.\ref{profile}). Discrepancies between the two
profiles are due, inside 2.5$\arcsec$ to the above-mentioned
diffraction spikes, and outside 3.6$\arcsec$ to an imperfect
elimination of the blurred light from the stellar companions \hdb\ and
\hdc.

We now focus on the Southern profile which shows more clearly the
annular shape of the disk.  Both inside and outside the peak at
325\,AU, the surface brightness steeply decreases (FWHM $\sim$
150\,AU). The decline becomes smoother further than 420\,AU. More
precisely, Table \ref{slopes} gives the steepness indexes which match
the surface brightness for different ranges of distance from the star.
The change of slope around $4.2\arcsec$ has to be confirmed.  Indeed,
this distance corresponds to the position of a different detector
matrix, also we do not exclude that shading may have caused this
effect \citep{stsci}.  Nevertheless, the disk is positively detected
at least up to $6\arcsec$.

\begin{table}[!h] 
\begin{center}
\caption{\label{slopes} Surface brightness (SB) steepness indexes $\alpha$
  (SB $\propto r^{\alpha}$) and corresponding SB ranges in
  mag.arcsec$^{-2}$ for the Southern extension. The zero point flux in
  F160W filter given by the STScI is 1113\,Jy.}
\begin{tabular}[h]{ccc}
\hline
Major Axis & Steepness & SB Range\\
Radial Range & Index $\alpha$ & in mag.arcsec$^{-2}$\\
\hline
$2.7\arcsec\rightarrow 3.2\arcsec$ & $+1.98\pm 0.14$ & $17.3\rightarrow 16.9$ \\
$3.3\arcsec\rightarrow 3.6\arcsec$ & $-2.32\pm 0.16$ & $16.9\rightarrow 17.2$ \\
$3.6\arcsec\rightarrow 4.2\arcsec$ & $-6.87\pm 0.14$ & $17.2\rightarrow 18.3$ \\
$4.2\arcsec\rightarrow 6.0\arcsec$ & $-2.75\pm 0.20$ & $18.3\rightarrow 19.5$ \\
\hline
\end{tabular}
\end{center}
\end{table}
We performed the photometry of the disk on elliptic contours with
semi-major axis between $2\arcsec$ and $9\arcsec$. We find a total
scattered flux density of $4.5\pm 0.5$\,mJy ($\sim 13.5$ mag), which
must be considered as a lower value since the scattered light below
the spider diffraction patterns is not taken into account.  This
corresponds to a ratio of scattered to stellar flux at 1.6\,$\mu$m of
about $2.2\times 10^{-3}$.
%
\section{Discussion}
\subsection{Disk properties}
\label{discussion}
Assuming an optically thin ring and an inclination from edge-on of
$35\degr$, we reproduce the main shape of the surface brightness along
the major axis of the disk with an annulus peaked at 330\,AU from the
star and a radial surface dust distribution proportional to $r^{4}$
inside the peak and to $r^{-6.8}$ outside. This profile does not
depend on the exact anisotropic scattering properties of the grains
because it is measured along the major axis of the disk, \textit{i.e.}
where the scattering angle always close to $90\degr$.  For simplicity,
we have therefore assumed that grains scatter isotropically. The
predicted surface brightness is superimposed in Figure \ref{profile}
on the observed one.

The dust population derived, assumed to be made of amorphous fluffy
grains as described in \citet{aug99} larger than about a half
micrometer, fits quite well the 20--100\,$\mu$m SED but does not
explain the shorter wavelength data.  Emission features at
7.7\,$\mu$m, 8.6\,$\mu$m and 11.3\,$\mu$m, tentatively attributed to
aromatic molecules (e.g.  Polycyclic Aromatic Hydrocarbons), have been
detected by \citet{syl96a}. The present model can not reproduce these
features. Anyway, the presence of dust closer to the star is required
to reproduce at least the 10\,$\mu$m continuum. This second population
is expected to be too close to the star (typically inside the first
hundred AU) to be detectable in the present data.  Such hot grains
probably contribute to the $20\,\mu$m SED and may also be responsible
for all or part of the 12.5\,$\mu$m and 17.9\,$\mu$m extended
emissions (0.75$\arcsec$ (75\,AU) in radius) resolved by
\citet{sil98}.
More data and modeling are needed to confirm that issue.
\begin{figure}[tbp]
\begin{center}
\includegraphics[angle=90,origin=bl,width=0.5\textwidth]{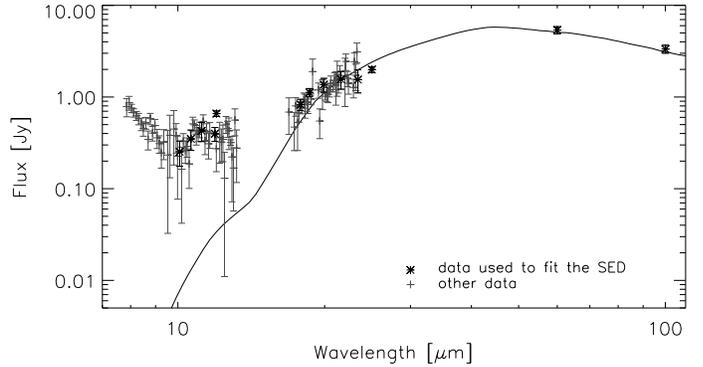}
\caption[]{Best fit of the 8--100\,$\mu$m SED
  (IRAS and \citet{syl96a} measurements) taking into account the fact
  that the disk shapes an annulus.
  Grains are assumed to be made of silicate, organic refractories and
  H$_2$O ice and follow a collisional size distribution ($\propto
  a^{-3.5}$). The long wavelength continuum is well reproduced whereas
  the 10\,$\mu$m predicted flux densities are about one order of
  magnitude smaller than the measured ones. This model suggests that
  an additional population of hot grains closer to the star is needed
  to account for the short wavelength excesses.}
\label{SED}
\end{center}
\end{figure}
\subsection{Comparison with \hra}
It is particularly instructive to further compare \hda\ to \hra:
\begin{itemize}
\item both stars exhibit a \cs\ ring, but the \hda\ annulus is
  about 9--10 times wider than the \hra\ one,
\item two dust populations are required to fit both full SEDs
  \citep{koe98,aug99},
\item the inner edge of the \hda\ disk suggests a truncation process
  as already proposed for \hra,
\item the outer disk distribution (further $\sim$325\,AU) seems
  steeper than the spatial distribution of dust supplied by colliding
  or evaporating bodies \citep{lec96} as for \hra. A perturbing body
  as a source for this outer truncation is possible. Nevertheless, no
  massive perturbing body is detected so far unlike for \hra. 
\end{itemize}
Most of the above remarks regarding the properties and the dynamics of
the \hda\ \cs\ disk will be further investigated in a forthcoming
paper.

\end{document}